\documentclass[12pt]{article}
\pdfoutput=1

\usepackage[toc,page]{appendix}
\usepackage{slashed}
\usepackage[sumlimits,intlimits,namelimits]{amsmath} 
\usepackage[english]{babel}
\usepackage{graphicx}
\usepackage{hyperref}
\usepackage{cite}
\usepackage{soul}

\newcommand{\ba}{\begin{eqnarray}}
\newcommand{\ea}{\end{eqnarray}}

\newcommand{\ice}[1]{\relax}

\usepackage{color}
\usepackage{amsmath}
\usepackage{booktabs} 
\usepackage{subfigure}
\DeclareMathOperator{\Tr}{Tr}

\begin{document}

\begin{titlepage}
  
\begin{flushright}
SI-HEP-2025-05\\
SFB-257-P3H-25-019\\
PSI-PR-25-06
\end{flushright}

\vspace{0.12cm}
\begin{center}
	  { \Large\bf
	  QCD corrections for subleading powers in \boldmath $1/m_b$ \unboldmath \\ [2mm]
	  for the nonleptonic $b\rightarrow c\bar c s$ transition\\
}
\end{center}
%
\vspace{0.35cm}
\begin{center}
  {\sc Thomas Mannel${}^1$}, {\sc Daniel Moreno${}^2$},
  and {\sc Alexei A. Pivovarov${}^1$} \\[0.2cm]
  {\sf ${}^1$Center for Particle Physics Siegen,
    Theoretische Physik 1, Universit\"at Siegen\\ 57068 Siegen, Germany} \\[1mm] 
  {\sf ${}^2$Laboratory for Particle Physics, PSI Center for Neutron and Muon Sciences,
Forschungsstrasse 111, 5232 Villigen PSI, Switzerland}
\end{center}

\vspace{0.48cm}
\begin{abstract}\noindent
We compute next-to-leading order QCD corrections to the $1/m_b^2$ terms of the Heavy Quark Expansion for the contribution to inclusive nonleptonic bottom hadron decays induced by the charged-current operators mediating the transitions $b \to c \bar c q$ ($q = s,d$). Their contributions to lifetimes are Cabibbo favoured and have large Wilson coefficents, but have not been computed before, since the presence of two heavy quarks
in the final state makes this calcuation technically demanding.   
It finally adds the last missing
piece of the full NLO terms at $1/m_b^2$ in the heavy quark expansion of $B$-hadron lifetimes.
We obtain analytical results with full dependence on the final-state charm quark mass,
which we provide in the supplemental file \textit{"coefbccs.m"}.
\end{abstract}
\end{titlepage}

\section{Introduction} 
\label{sec:Intro} 
The Heavy Quark Expansion (HQE)~\cite{Chay:1990da,Bigi:1992su,Bigi:1993fe,Blok:1993va,Manohar:1993qn}
has become the standard tool to calculate inclusive decays of 
heavy hadrons~\cite{Lenz:2014jha}, i.e. hadrons containing one heavy quark. The most inclusive quantities are the
lifetimes of heavy hadrons, and the HQE seems to work quite well for bottom hadrons, where 
the predictions of the HQE provide a satisfactory picture~\cite{Lenz:2022rbq,Gratrex:2023pfn,Albrecht:2024oyn}. However,
it is well known that spectator effects, which appear first at order $1/m_Q^3$, are enhanced 
by phase space factors of $16 \pi^2$. While these contributions are still 
small for bottom hadrons ($16 \pi^2 \Lambda_{\rm QCD}^3 / m_b^3 \sim 0.2)$, 
they become sizable for the case of charmed hadrons
($16 \pi^2 \Lambda_{\rm QCD}^3 / m_c^3 \sim 5)$, making the reliability
of the HQE for charmed hadrons questionable~\cite{King:2021xqp,Gratrex:2022xpm,Dulibic:2023jeu}.

From the experimental side, the lifetimes of the ground-state bottom and 
charm hadrons have been measured very precisely, in many cases with precision well below
one percent. However, theoretical predictions for total rates suffer from larger uncertainties
due to the fifth power of the quark mass appearing at leading order in the HQE, requiring 
calculations in a mass scheme, for which a precise input can be obtained. On the other hand, 
lifetime ratios can be computed in the HQE with less uncertainties, since the factor $m_Q^5$
cancels in the ratio. The benchmark for the theoretical calculations are the current 
measurements~\cite{HFLAV:2022esi}
\begin{equation}
	\frac{\tau(B_s)}{\tau(B_d)}\bigg|^{{\mbox{\scriptsize exp}}} = 0.998 \pm 0.005\,,
	\;\;\;\;
	\frac{\tau(B^+)}{\tau(B_d)}\bigg|^{{\scriptsize\mbox{exp}}} = 1.076 \pm 0.004\,,
	\;\;\;\;
	\frac{\tau(\Lambda_b)}{\tau(B_d)}\bigg|^{{\scriptsize\mbox{exp}}} = 0.969 \pm 0.006\,,
\end{equation} 
which have a sub-percent precision in the meantime. 

Therefore, calculations of higher-order corrections in $1/m_Q$ as well as in $\alpha_s$ for 
lifetimes and lifetime ratios are continuously improved, in particular 
for the nonleptonic contributions of the effective Hamiltonian. The leading
term of the HQE is always the partonic contribution, i.e. the decay of a ``free'' (heavy)
quark. The leading order (LO) and next-to-leading order (NLO) contributions to the partonic rate are known since
long~\cite{Altarelli:1980fi,Buchalla:1992gc,Bagan:1994zd,Czarnecki:2005vr,Krinner:2013cja}.
The next-to-next-to-leading order (NNLO) partonic rate has been evaluated only recently~\cite{Egner:2024azu}, which has
removed the large uncertainty in the leading term due to the dependence on the 
renormalization scale~\cite{Egner:2024lay}.

At subleading powers, the coefficients of the
$1/m_b^2$ corrections have been computed at LO~\cite{Bigi:1992su,Blok:1992hw,Blok:1992he} 
and at NLO in the case of massless quarks in the final state~\cite{Mannel:2023zei}. The 
coefficients of the $1/m_b^3$ corrections are known at LO for the two-quark
operators~\cite{Lenz:2020oce,Mannel:2020fts,Moreno:2020rmk}
and at NLO for the four-quark operators~\cite{Beneke:2002rj,Franco:2002fc}. Finally, 
the coefficients of the $1/m_b^4$
corrections are known at LO for the four-quark operators~\cite{Gabbiani:2004tp}. To this end, 
the present calculation completes the NLO contributions at $1/m_Q^2$ for the case of the
bottom quark. 

We point out that the $\alpha_s$ corrections to the $1/m_Q^2$ suppressed terms
for the CKM favoured $b\rightarrow c \bar u d$ transition calculated in~\cite{Mannel:2024uar}
were found to be very important, since they overcome the leading contribution 
due to accidental numerical cancellations. Previous to this calculation, 
even the sign of the coefficient of the chromomagnetic moment contribution was unclear. 
In light of the relevance of these corrections we address in this paper the
calculation of the $\alpha_s$ corrections to the chromomagnetic coefficient
for the other CKM-favoured transition $b\rightarrow c\bar c s$.

Thus we follow-up our previous works~\cite{Mannel:2023zei,Mannel:2024uar} where
$\alpha_s$ corrections to the perturbative coefficients of the first power suppressed terms
in the HQE of inclusive nonleptonic decays of
heavy hadrons was considered for the case of massless final-state quarks ($b\rightarrow u \bar u d$ and $c\rightarrow s \bar d u$)
and one massive final-state quark ($b\rightarrow c \bar u d$).
To this end, the calculation presented here completes the NLO contributions at $1/m_Q^2$ for the case of the
bottom quark. 
  
We obtain analytical results with full dependence on the final-state quark mass $m_c$
in terms of multiple polylogarithms (MPLs) defined recursively via iterated integrals~\cite{goncharov2011multiplepolylogarithmscyclotomymodular,goncharov2001multiplepolylogarithmsmixedtate}.
The master integrals required for the calculation were obtained in~\cite{Egner:2024azu}.
Since results are quite lengthy we provide them in the supplemental material \textit{``coefbccs.nb''}
as a Mathematica notebook.

The paper is organized as follows.
In section~\ref{sec:LeffEW} we discuss the operators of the weak effective Lagrangian that are relevant for
the calculation, the renormalization scheme, including the choice of evanescent operators, and the scheme used for $\gamma_5$.
We provide the definitions within the HQE in section~\ref{sec:hqerate}.
In section~\ref{sec:outline} we outline the computation and discuss the impact of the results in section~\ref{sec:res}.

\section{The Weak Effective Lagrangian}
\label{sec:LeffEW}
We consider the inclusive weak decay of a bottom hadron induced by the quark transitions  
$b\rightarrow c\bar c q$ with $q =s$ or $d$, neglecting the masses of the $u$, $d$ and $s$ 
quarks. The bottom and the charm-quark masses are kept in the calculation, assuming
$\rho = m_c^2 / m_b^2$ to be of order unity. The relevant $\Delta B=1$ effective weak Lagrangian
is given by   
%
%
%
%
\begin{equation}
 \mathcal{L}_{{\scriptsize\mbox{eff}}} = - 2\sqrt{2} G_F V_{c q} V_{cb}^* (C_1 \mathcal{O}_1 + C_2 \mathcal{O}_2)
 + \mbox{h.c}\,,
 \label{eq:FermiLagr}
\end{equation}
which in the standard basis contains color singlet and color rearranged operators
\begin{eqnarray}
\label{basis}
 \mathcal{O}_1 &=& (\bar b^i \gamma^\mu P_L c^j)(\bar c^j \gamma_\mu P_L q^i)\,,
 \\
 \mathcal{O}_2 &=& (\bar b^i \gamma^\mu P_L c^i)(\bar c^j \gamma_\mu P_L q^j)\,,
\end{eqnarray}
where $G_F$ is the Fermi constant, $V_{qq'}$ are CKM matrix elements, $C_{1,2}$ are the Wilson
coefficients, $P_L = (1-\gamma_5)/2$ is the left-handed projector and $(i,j)$ are color indices.
Note that we have neglected QCD and electroweak penguin operators, which have small Wilson 
coefficients.


For the renormalization it is nevertheless convenient to work in an operator basis for which the anomalous dimension matrix 
becomes diagonal
%
\begin{equation}
\mathcal{O}_\pm = \frac{1}{2} (\mathcal{O}_2 \pm \mathcal{O}_1)  \quad \mbox{and} \quad  C_\pm = C_2 \pm C_1 \, ,
\end{equation}
with renormalization constants $Z_{\pm}$ in the $\overline{\mbox{MS}}$ renormalization scheme given by
\begin{eqnarray}
 C_{\pm\,,B} = Z_{\pm} C_\pm \,,\quad\quad
 Z_{\pm} = 1 - \frac{3}{N_c} (1 \mp N_c) \frac{\alpha_s(\mu)}{4\pi}\frac{1}{\epsilon}\,,
\end{eqnarray}
where the subscript $B$ indicates bare quantities and those with omitted subscript indicate renormalized quantities.
The constant $N_c=3$ is the number of colors.

The presence of $\gamma_5$ in the operators $\mathcal{O}_{\pm}$ requires to specify its treatment in 
dimensional regularization.   
It has become a common choice to work in naive dimensional regularization (NDR), i.~e. with anticommuting 
$\gamma_5$, together with a choice of the evanescent operators that preserves Fierz symmetry at a given
order in the $\alpha_s$ expansion~\cite{Buras:1989xd,Dugan:1990df,Herrlich:1994kh}
(see also \cite{Egner:2024azu} for a comprehensive description). 


At NLO, preservation of Fierz symmetry is achieved by choosing the following set of evanescent operators~\cite{Buras:1989xd}
\begin{eqnarray}
\label{evanescent}
 E_1 &=& (\bar b^i \gamma^\mu \gamma^\nu \gamma^\alpha P_L c^j)(\bar c^j \gamma_\mu \gamma_\nu \gamma_\alpha P_L q^i) - (16-4\epsilon)\mathcal{O}_1\,,
 \\
 E_2 &=& (\bar b^i \gamma^\mu \gamma^\nu \gamma^\alpha P_L c^i)(\bar c^j \gamma_\mu \gamma_\nu \gamma_\alpha P_L q^j) - (16-4\epsilon)\mathcal{O}_2\,.
\end{eqnarray}
In this scheme, the Wilson coefficients $C_{\pm}$ of the physical operators are known at NNLO~\cite{Gorbahn:2004my}
and at next-to-next-to-next-to-leading logarithmic (NNLL) precision~\cite{Altarelli:1980fi,Buras:1989xd,Bagan:1994zd,Gorbahn:2004my}.
The renormalization group improved coefficients at next-to-leading logarithmic (NLL) precision are given by
\begin{eqnarray}
	C_{\pm}(\mu) = L_{\pm}(\mu)\bigg[1 + \frac{\alpha_s(M_W) - \alpha_s(\mu)}{4\pi}R_{\pm} + \frac{\alpha_s(\mu)}{4\pi}B_{\pm}\bigg]\,.
	\label{CpmNLO}
\end{eqnarray}
These coefficients have been calculated at the renormalization scale $\mu = M_W$ and then evolved down to scales $\mu\sim m_b \ll M_W$.
The term proportional to $B_{\pm}$ encodes the scheme-dependence of the
coefficients which eventually cancels the scheme-dependence of the correlators. Explicitly, the quantities in $C_\pm$ read
\begin{eqnarray}
	R_{+} &=& \frac{10863 - 1278n_f + 80n_f^2}{6(33-2n_f)^2}\,,\quad\quad\quad\quad
	R_{-} = - \frac{15021 - 1530n_f + 80n_f^2}{3(33-2n_f)^2}\,,
	\nonumber
	\\
	B_{\pm} &=& -\frac{1}{2N_c} B (1 \mp N_c) \,, \quad\quad\quad\quad
	L_{\pm}(\mu) = \bigg(\frac{\alpha_s(M_W)}{\alpha_s(\mu)}\bigg)^{-\frac{3}{\beta_0 N_c}(1 \mp N_c)}\,,
	\label{RpmBpm}
\end{eqnarray}
where $n_f$ is the number of light flavours, $B=11$ in NDR and $\beta_0 = \frac{11}{3}N_c - \frac{2}{3} n_f$.

\section{HQE for nonleptonic decays of $B$-hadrons up to $(\Lambda_{\rm QCD}/m_b)^2$}
\label{sec:hqerate}

We set up the HQE by following our previous 
works~\cite{Mannel:2021zzr,Moreno:2022goo,Mannel:2023zei,Moreno:2024bgq,Mannel:2024uar}
and adopting the same definitions.
We calculate the inclusive nonleptonic decay width $\Gamma$ of a $B$-hadron by using the optical theorem

%
\begin{equation} 
	\Gamma (B \to X) =
	\frac{1}{M_B} \text{Im } \langle B(p_B)|i\, \int d^4 x\, T\{ {\cal L}_{\rm eff} (x)  {\cal L}_{\rm eff} (0) \}
	|B(p_B) \rangle \,,\quad\quad
	\label{gamma}
\end{equation} 
where $p_B = M_B v$ is the momentum of the $B$-hadron state $|B\rangle$ with mass $M_B$ and velocity $v$.
The heavy quark momentum is decomposed as $p_b = m_b v + k$, where $k \sim \Lambda_{\rm QCD} \ll m_b$
is a small residual momentum of the heavy quark inside the hadron with components of order
$\Lambda_{\rm QCD}$. The HQE is set up as an expansion in $k$, which amounts in an expansion
of Eq.~(\ref{gamma}) in $\Lambda_{\rm QCD}/m_b$  
\begin{equation}
	\label{hqewidth}
	\Gamma(B\rightarrow X) = \Gamma_0 |V_{cq}|^2 |V_{cb}|^2
	\bigg(  C_0
	- C_{\mu_\pi}\frac{\mu_\pi^2}{2m_b^2}
	+ C_{\mu_G}\frac{\mu_G^2}{2m_b^2}
	\bigg)\,,\quad\quad \Gamma_0= \frac{G_F^2 m_b^5}{192 \pi^3}\,,
\end{equation}
where $\mu_\pi^2,\, \mu_G^2 \sim \Lambda_{\rm QCD}^2$ are HQE parameters representing the nonperturbaive input.

Note that Eq.~(\ref{hqewidth}) factorizes perturbative contributions, encoded in the
Wilson coefficients $C_i$, from the non-perturbative input $\mu_\pi^2$ and $\mu_G^2$.
The Wilson coefficients $C_i$ ($i=0,\,\mu_\pi,\,\mu_G$) have a perturbative expansion in $\alpha_s(\mu)$, with
$\mu\sim m_b$, and depend only on the ratio of two scales $\rho = m_c^2/m_b^2$.
The nonperturbative HQE parameters are defined from the local HQET operators (up to dimension five)
\begin{eqnarray}
	\mathcal{O}_0 &=& \bar h_v h_v\,,\quad\quad\quad\quad\quad\quad\quad
	\mathcal{O}_v = \bar h_v v\cdot iD h_v\,,
	\nonumber
	\\
    \mathcal{O}_\pi &=& \bar h_v (iD_\perp)^2 h_v\,,\quad\quad\quad\quad
	\mathcal{O}_G = \frac{1}{2}\bar h_v [\gamma^\mu, \gamma^\nu] iD_{\perp\,\mu} iD_{\perp\,\nu}  h_v\,,\quad\quad\quad\quad
\label{eq:OpList}
\end{eqnarray}
as their forward matrix elements~\cite{Mannel:2018mqv}
\begin{eqnarray}
	\langle B(p_B)\lvert \bar b \slashed v b \lvert B(p_B)\rangle &=& 2M_B\,,  \\
	- \langle B(p_B)\lvert \mathcal{O}_\pi \lvert B(p_B)\rangle &=& 2M_B \mu_\pi^2\,, \\
	c_F(\mu)\langle B(p_B)\lvert \mathcal{O}_G \lvert B(p_B)\rangle
	&=& 2M_B \mu_G^2\,,
\end{eqnarray}
where $h_v$ is the static HQET field, $D_\mu = \partial_\mu - ig_s A_\mu^a T^a$ is the covariant derivative of QCD, 
and perpendicular quantities  are defined as $a^\mu_\perp = a^\mu -v^\mu (v\cdot a)$.
Note that in Eq.~(\ref{hqewidth}) only $\mu_\pi^2$ and $\mu_G^2$ appear. All other forward matrix elements of the opertators in (\ref{eq:OpList}) can be related to $\mu_\pi^2$ and $\mu_G^2$.

Finally, following our previous works, we include
the chromomagnetic operator coefficient of the HQET Lagrangian
\begin{eqnarray}
	c_F(\mu) &=& 1 + \frac{\alpha_s(\mu)}{2\pi}\bigg[ \frac{N_c^2-1}{2N_c} + N_c\bigg(1 + \ln\bigg(\frac{\mu}{m_b}\bigg)\bigg) \bigg]
\end{eqnarray}
into the definition of $\mu_G^2$, relating $\mu_G^2$ directly
to the mass splitting of the ground-state $B$ mesons
\begin{equation}
\mu_G^2 = \frac{3}{4} (M_{B^*}^2 - M_B^2)\,.
\end{equation}

\section{Outline of the calculation}
\label{sec:outline}
The present calculation follows a very similar scheme as the one in our previous
papers~\cite{Mannel:2021zzr,Moreno:2022goo,Mannel:2023zei,Moreno:2024bgq,Mannel:2024uar}. 
%
%
In the first step, we match the time ordered product in Eq.~(\ref{gamma}) to HQET, using the
operator basis (\ref{eq:OpList})
\begin{eqnarray}
	&& 
    \text{Im } i\, \int d^4 x\, T\{ {\cal L}_{\rm eff} (x)  {\cal L}_{\rm eff} (0) \} 
    \nonumber
	\\
	&=&
    \Gamma_0 |V_{cq}|^2 |V_{cb}|^2   
	\bigg( C_0 \mathcal{O}_0
	+ C_v \frac{\mathcal{O}_v}{m_b}
	+ C_\pi \frac{\mathcal{O}_\pi}{2m_b^2}
	+ C_G \frac{\mathcal{O}_G}{2m_b^2} + \mathcal{O}\left(\frac{\Lambda_{\rm QCD}^3}{m_b^3}\right)
	\bigg)  \,,
	 \label{eq:trans_operator}
\end{eqnarray}
%
where the perturbative coefficients $C_i$ ($i=0,\,v,\,\pi,\,G$) depend only on $\rho$.
Note that (\ref{eq:trans_operator}) is an operator relation with universal coefficients  $C_i$ ($i=0,\,v,\,\pi,\,G$).
This relation thus can be evaluated for any kind of external state. To obtain the total rate, we take a forward matrix element of the
full-QCD hadron state $|B(p_B) \rangle$, while the coefficients $C_i$ ($i=0,\,v,\,\pi,\,G$) 
are computed by a matching to matrix elements with quarks and gluons. 

In a second step, we trade the operator $\mathcal{O}_0$ for the local
operator $\bar b \slashed v b$ defined in full QCD, which is related to the 
conserved current $\bar b \gamma_\mu b$. Its forward matrix element is thus the 
$b$ quark number of the state, which is for a bottom hadron exactly unity, to all 
orders in $\alpha_s$. This trade is convenient, since it shifts the appearance of 
nonperturbative parameters to the order $\Lambda_{\rm QCD}^2/m_b^2$.
In fact,  matching of the QCD current to HQET yields
\begin{equation}
	\bar b \slashed v b = \mathcal{O}_0 + \tilde{C}_v \frac{\mathcal{O}_v}{m_b} + \tilde C_\pi \frac{\mathcal{O}_\pi}{2m_b^2}
	+ \tilde C_G \frac{\mathcal{O}_G}{2m_b^2}  \,,
	\label{hqebvb}
\end{equation}
where we include only terms up to order $\Lambda_{\rm QCD}^2/m_b^2$. 
After using the equation of motion of the HQET Lagrangian for the operator $\mathcal{O}_v$
\begin{eqnarray}
	\mathcal{O}_v =
	- \frac{1}{2m_b} (\mathcal{O}_\pi + c_F(\mu)\mathcal{O}_G) + \mathcal{O}\left(\frac{\Lambda_{\rm QCD}^2}{m_b^2}\right)\,,
	\label{LHQET}
\end{eqnarray}
we finally obtain, up to order $\Lambda_{\rm QCD}^2/m_b^2$,
\begin{eqnarray}
	\label{hqedifwidth}
	\Gamma(B \to X)
	&=& \Gamma_0 |V_{cq}|^2 |V_{cb}|^2
	\bigg[ C_0 \bigg( 1
	- \frac{\bar{C}_\pi - \bar{C}_v }{C_0}\frac{\mu_\pi^2}{2m_b^2}\bigg)
	+ \bigg(\frac{\bar{C}_G}{c_F(\mu)} -  \bar{C}_v \bigg)\frac{\mu_G^2}{2m_b^2} \bigg] \, ,
\end{eqnarray}
with $\bar{C}_i\equiv C_i - C_0 \tilde{C}_i$ ($i=v,\,\pi,\,G$), from which we can read off the 
coefficients from eq.~(\ref{hqewidth})
\begin{eqnarray}
C_{\mu_\pi} = \bar{C}_\pi - \bar{C}_v = C_0\,,\quad\quad\quad\quad\quad\quad
C_{\mu_G} = \frac{\bar{C}_G}{c_F(\mu)} -  \bar{C}_v\,,
\end{eqnarray}
where the last equality in the first equation is a consequence of reparametrization invariance. Therefore, the computation
reduces to the one of $C_0$ and $C_{\mu_G}$ together with the auxiliar coefficient $\bar{C}_v$.

\begin{figure}[!htb]
	\centering
	\includegraphics[width=1.0\textwidth]{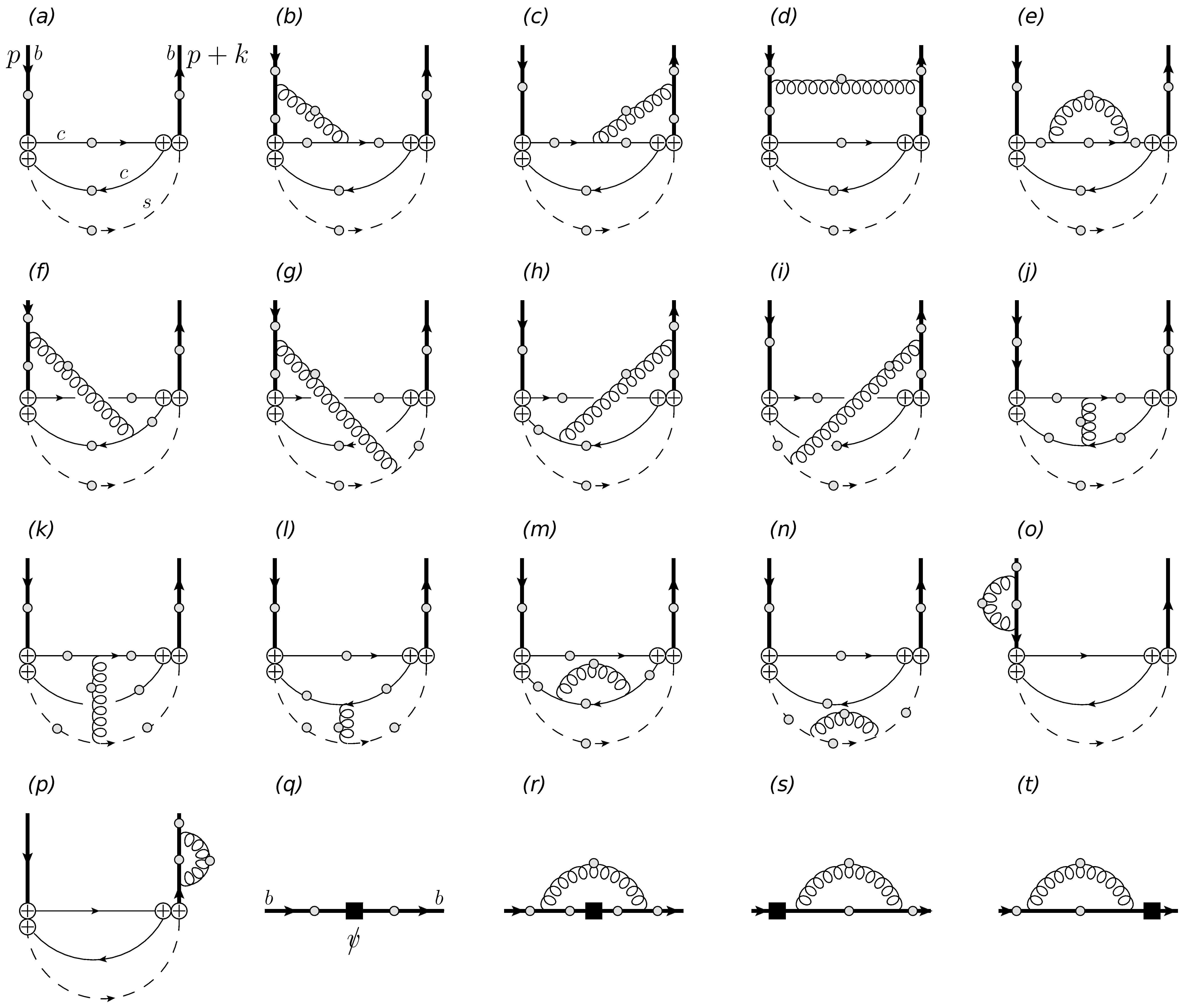}
      \caption{Feynman diagrams (a)-(t) contributing to the coefficients $C_0$, $\bar C_v$ and $\bar C_G$ of the HQE of the nonleptonic decay width up to NLO. The incoming heavy quark carries momentum $p$, with $p^2=m_b^2$. Grey dots stands for possible gluon insertions with incoming momentum $k\sim \Lambda_{\rm QCD}$. The black box vertex stands for $\slashed v$ insertions.
      The diagrams contributing to power corrections are obtained after taking into account all possible one-gluon insertions.
        Four-fermion vertices correspond $\mathcal{O}_\pm$ insertions of $\mathcal{L}_{\rm eff}$ after applying an appropriate Fierz transformation.}
                \label{NLPdiagrams}
\end{figure}

The computation of the matching coefficients is analogous to the one carried out
in our previous works~\cite{Mannel:2023zei,Mannel:2024uar} (see also \cite{Moreno:2024bgq,Moreno:2022goo,Mannel:2021zzr}),
with the only difference that we have now one massive line of mass $m_b$, two massive lines
of mass $m_c$ and one massless line. Therefore we will proceed exactly as described in 
our previous works,
where we take the corresponding amplitude given by the diagrams shown in fig.~\ref{NLPdiagrams},
expand in the small momentum $k$ up to the needed order, and project it to the contribution of the desired operator.

The coefficient $C_0$ is given by the projection
\begin{eqnarray}
 C_0 \sim \Tr(\mathcal{A}_1 P_{+}),
\end{eqnarray}
of the $b(p)\rightarrow b(p)$ scattering amplitude $\mathcal{A}_1$ from diagrams (a)-(p) 
of fig.~\ref{NLPdiagrams} with $p^2=m_b^2$ and
$P_{+} =(1+\slashed v)/2$.

The coefficients $\bar C_v$ and $\bar C_G$ are given by the projections
\begin{eqnarray}
  \bar C_v \sim v_\lambda \Tr( \mathcal{A}_0^\lambda P_{+})\,, \quad\quad\quad\quad
  \bar C_G \sim \frac{1}{(D-2)(D-1)} \Tr( \mathcal{A}_k^{\lambda \alpha} P_{+} [\gamma_{\perp\,\alpha},\gamma_{\perp\,\lambda}] P_{+})
\end{eqnarray}
of the $b(p)\rightarrow b(p+k)g(-k)$ scattering amplitude $\mathcal{A}_2^\lambda = \mathcal{A}_0^\lambda + \mathcal{A}_k^{\lambda \alpha} k_\alpha + \mathcal{O}(k^2)$ from diagrams (a)-(t) of fig.~\ref{NLPdiagrams} expanded up to linear order in the small momentum $k$ with $p^2=m_b^2$.
The scattering amplitudes are computed in the Feynman gauge and by using the background field method.

As already discussed in sec.~\ref{sec:LeffEW} we use NDR in $D=4-2\epsilon$ space-time dimensions with the scheme
of evanescent operators preserving Fierz symmetry to order $\alpha_s$. This choice allows us to apply Fierz symmetry when making insertions of the $\mathcal{O}_{1,2}$ operators such that the corresponding Feynman diagrams contain a single open fermionic line, avoiding in this way inconsistencies due to the use of anticommuting $\gamma_5$ in dimensional regularization. Specifically, we chose
to apply always the Fierz transformation to the second insertion. This results into a scheme dependence of the correlators
which cancels against the scheme dependence of the Wilson coefficient of the weak effective Lagrangian, rendering the
Wilson coefficients of the HQE scheme-independent.
Also note that up to NLO only correlators of physical operators need to be computed.

For the renormalization of the bottom and charm quarks we use the on-shell scheme. For $\alpha_s$ and the HQET Lagrangian
we use the $\overline{\mbox{MS}}$ scheme. The renormalization constants can be taken e.~g. from~\cite{Moreno:2022goo}.

Note from Fig.~[\ref{NLPdiagrams}] that we need to compute the imaginary part of two-loop and three-loop diagrams
at LO and NLO, respectively. This involves cutting two massive propagators of mass $m_c$ and one or two massless propagators.
For the calculation we employ a number of tools which provide a high level of automation.
For the generation of the (projected) amplitude we use an in-house Mathematica code together with
{\tt Tracer}~\cite{Jamin:1991dp} and {\tt ColorMath}~\cite{Sjodahl:2012nk} to handle the Dirac and color algebra, respectively.
Once the amplitude is obtained we implement the integration-by-parts reduction (IBP) by using {\tt LiteRed}~\cite{Lee:2012cn,Lee:2013mka}, which allows to write the amplitudes as a combination of 2 master integrals at LO and 17 master integrals at NLO.
The required master integrals have been computed previously in ref.~\cite{Egner:2024azu} by using the method of
differential equations, in terms of MPLs, that can be defined recursively via the iterated integral ($n\geq 0$)
\begin{eqnarray}
 G(a_1,\ldots,a_n; z) = \int_0^z \frac{dt}{t-a_1}G(a_2,\ldots,a_n;t)\,,
\end{eqnarray}
where $a_i$, $z$ are complex variables and the length of the weight vector $\vec a = (a_1,\ldots,a_n)$ is called the
weight. In the case where all the $a_i$'s are zero we define

\begin{eqnarray}
 G(\underbrace{0,\ldots,0}_{n};z) = \frac{1}{n!}\ln^n z\,.
\end{eqnarray}
Therefore, our results are quoted in terms of MPLs, which are expressed as functions of the variable $x$
\begin{eqnarray}
 x = \frac{1-\sqrt{1-4\rho}}{2\sqrt{\rho}}\,,\quad\quad\quad\quad  \sqrt{\rho} = \frac{x}{1+x^2}\,,
\end{eqnarray}
which is used in ref.~\cite{Egner:2024azu} to rationalize the system of differential equations previous to its transformation
to canonical form, which eventually allows to express the master integrals in terms of MPLs.
The phase space of the decay process allows to vary $\rho$ and $x$ in the range $0<\rho<1/4$ and $0<x<1$, respectively.

Indeed, it turns out that for the coefficients of the HQE, master integrals combine in such a way that only MPLs up to weight three appear, which can always be reduced to classical polylogarithms~\cite{Duhr:2019tlz}. We find that the final expressions for the coefficients contain at most trilogarithms.
We provide results in terms of both MPLs and polylogarithms. The former results are more compact and can be
evaluated numerically e.g. by using the {\tt Ginsh} command in {\tt PolyLogTools}~\cite{Duhr:2019tlz}. The later results are less compact but can be directly evaluated numerically with {\tt Mathematica} without the need of additional packages.
For the algebraic manipulation of MPLs we use both {\tt PolyLogTools} and {\tt HPL}~\cite{Maitre:2005uu}.

The results in both cases are too lengthy to be shown in the text. In the appendix~\ref{app:coefbccs}, we only provide the first few terms of the expanded results around $\rho=0$. The full analytical expressions for the coefficients of the HQE up to NLO and up to $1/m_b^2$
in the power expansion with full dependence on $m_c$ are provided in the supplemental file
\textit{"coefbccs.m"} in terms of MPLs and polylogarithms. The results consist of functions of the
ratio $\rho = m_c^2/m_b^2$ multiplying the coefficients of the weak effective Lagrangian
$C_{1,2}$. Both the coefficient functions and the Wilson coefficients $C_{1,2}$ depend
on the scheme used for $\gamma_5$ and the choice of evanescent operators in such a way the 
scheme dependence cancels for the coefficients of the HQE. In other words, the results provided 
in the file \textit{"coefbccs.m"} (and in the appendix) together with
Eqs.~(\ref{CpmNLO}) and (\ref{RpmBpm}) are scheme-independent quantities. 
Results are provided in terms of pole masses for the bottom and the charm quarks.

%

%


\section{Discussion of the results}
\label{sec:res}
In order to analyze the numerical impact of the $\alpha_s$ corrections to the $C_{\mu_G}$ coefficient, we
study its dependence on the renormalization scale $\mu$ and the parameter $\rho$. We
also analyze its impact on the inclusive nonleptonic $b\rightarrow c \bar c s$ decay width, which also requires to take into account the contribution of the already known $C_0$ coefficient at NLO. 

In the following plots, when we refer to the LO contribution, we include the calculation
of the correlators to LO, together with the resummation of logarithms in $C_{1,2}$ to leading logarithmic (LL) precision.
When we refer to the NLO contribution, we include the calculation of the correlators to NLO, together with the
resummation of logarithms in $C_{1,2}$ up to NLL precision.

We are not aiming at a full phenomenological analysis of the results, rather we want to estimate the 
overall effect of the $b \to c \bar{c} s$ NLO contribution to the subleading $1/m_b^2$ terms. To this
end, we pick some representative values for the parameters which we list in table~\ref{tab:par}.   
\begin{table}[h]
	\begin{center}
		\begin{tabular}{|c|c|c|c|}
			\hline
			Parameter & Numerical value & Parameter & Numerical value\\
			\hline
			$m_b$ & $4.7$ GeV  & $\mu_\pi^2$ & $0.5$ GeV$^2$\\
			$m_c$ & $1.6$ GeV &  $\mu_G^2$ &  $0.35$ GeV$^2$ \\
			$\rho=m_c^2/m_b^2$& $0.116$ & $M_W$ & $80.4$ GeV\\
			$\alpha_s(m_b)$ & $0.216$ & $\alpha_s(M_Z)$ & $0.120$\\
			\hline
		\end{tabular}
		\caption{Representative numerical values of the parameters entering the inclusive nonleptonic $b\rightarrow c\bar c s$ decay rate.}
		\label{tab:par}
	\end{center}
\end{table}

\begin{figure}[ht]
\centering
	\includegraphics[scale=0.67]{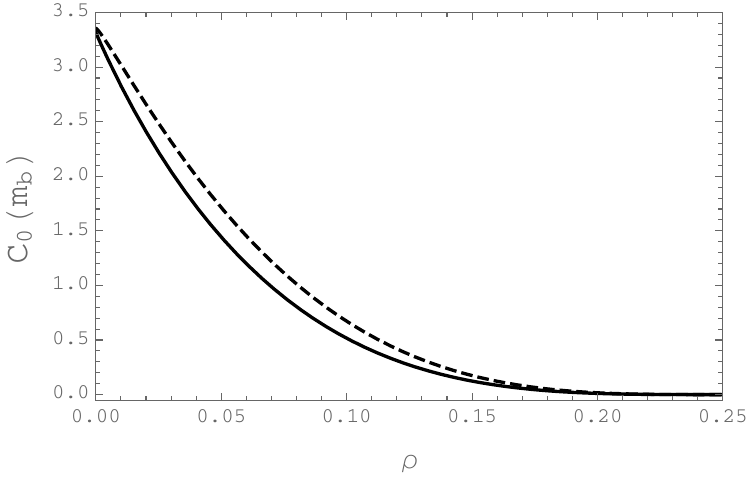}
\quad
\includegraphics[scale=0.67]{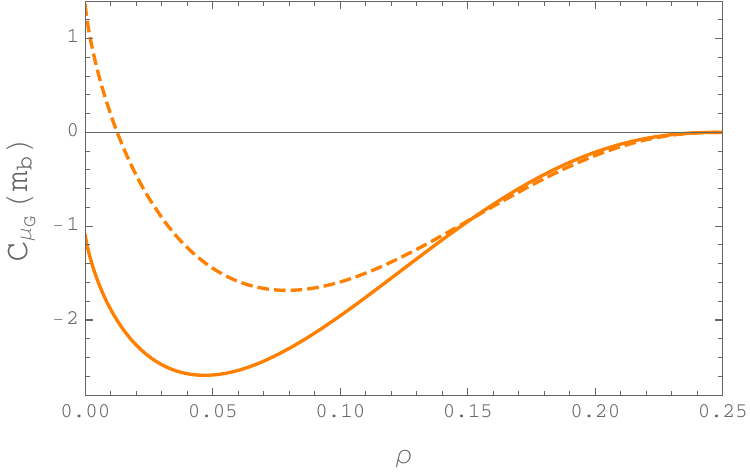}
\caption{$C_0$ (left) and $C_{\mu_G}$ (right) coefficients as a function of $\rho=m_c^2/m_b^2$ for $\mu=m_b$.
The solid lines show the coefficients at LO (with $C_{1,2}$ at LL), whereas the dashed lines include NLO corrections
(with $C_{1,2}$ at NLL).}
\label{fig:C0CGrho}
\end{figure}

In Fig.~\ref{fig:C0CGrho} we show the dependence of the $C_0$ and $C_{\mu_G}$ coefficients on the mass ratio $\rho$ at
LO and NLO, respectively.
We observe that the $C_0$ coefficient has a strong dependence on $\rho$ with a variation from
from $C_0(1/4)=0$ to $C_0(0)=3.35$. For the physical value of $\rho$ given in table~\ref{tab:par}
we find $C_0 = 0.46$, which is about a factor six smaller compared to the massless case due to the 
much smaller phase space available for this decay. The NLO corrections to $C_0$ also have some impact. We find,
for the representative value of $\rho$ given in table~\ref{tab:par}, a $34\%$ increase of the coefficient compared to its value at LO.

The $C_{\mu_G}$ coefficient strongly depends on the value of $\rho$ and has a very peculiar behaviour
shown in the the right panel of fig.~\ref{fig:C0CGrho}. While the NLO corrections are small for higher 
values of $\rho$, they can even overwhelm the LO contribution for smaller values of $\rho$. 
In particular, for $\rho=0$ the NLO coefficient has 
about the same value as the LO coefficient, but with opposite sign. 
This behaviour was already observed in~\cite{Mannel:2023zei,Mannel:2024uar} and is due to accidental 
numerical cancellations taking place at LO, enhancing the importance of the NLO corrections.
For the representative value of $\rho$ given in table~\ref{tab:par}, we find the NLO corrections
to increase the coefficient by $12\%$. However, we note that the $\rho$ dependence of the NLO contributions 
is steep, and for smaller values of $\rho$ the corrections become more important, e.g. for $\rho=0.05$
the NLO corrections increase the the coefficient by $44\%$. Thus the impact of the NLO corrections depend 
on the mass definitions used in a future full analysis. 

\begin{figure}[ht]
	\centering
	\includegraphics[scale=0.67]{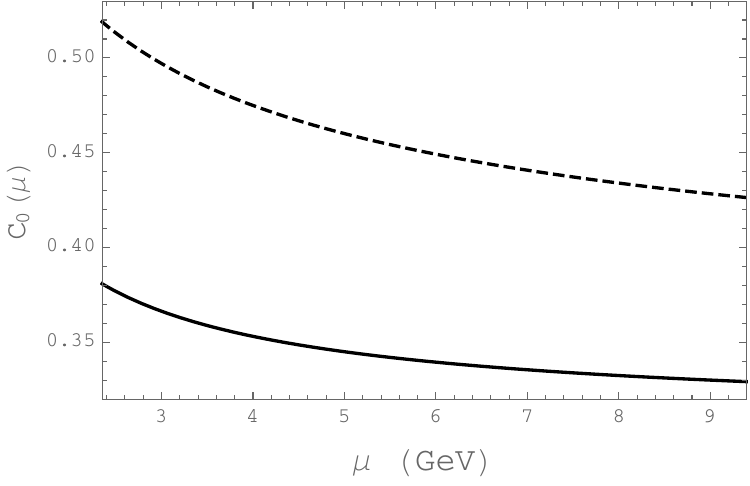}
	\quad
	\includegraphics[scale=0.67]{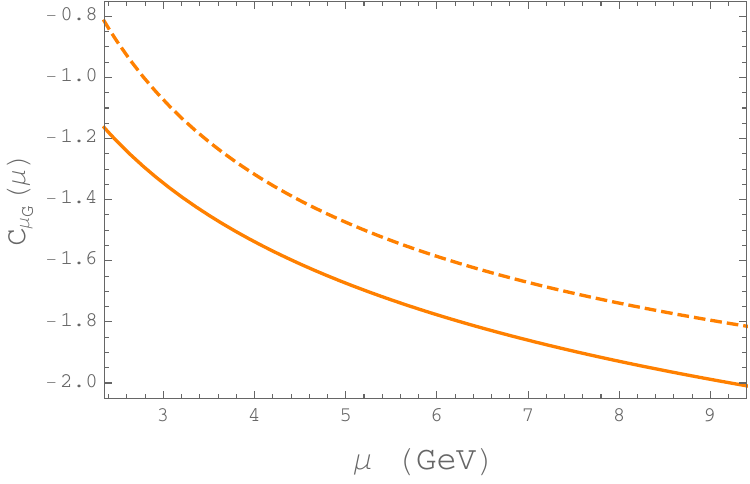}
	\caption{$C_0$ (left) and $C_{\mu_G}$ (right) as a function of the renormalization scale $\mu$ for $\rho=0.116$. The solid lines show the LO coefficients whereas the dashed lines include NLO corrections.}
	\label{fig:C0CGmu}
\end{figure}

In Fig.~\ref{fig:C0CGmu} we show the dependence of the $C_0$ and $C_{\mu_G}$ coefficients on the renormalization scale $\mu$ for our representative value $\rho=0.116$. The solid lines show the 
LO (i.e. LO computation of the correlator and LL for the Wilson coefficients $C_{1,2}$), while 
the dashed lines are the NLO results (i.e. NLO computation of the correlator and NLL for the 
Wilson coefficients $C_{1,2}$). We plot the coefficients in the range $m_b/2 \le \mu \le 2 m_b$ which 
is the typical one used to estimate the theory error of the coefficients. 

We observe a stronger $\mu$-dependence in the $C_{\mu_G}$ coefficient than in the $C_0$ coefficient leading to a higher scale uncertainty in $C_{\mu_G}$. However, overall the effect of the $\mu$ dependence in $C_{\mu_G}$ remains small for the total rate, since the chromomagnetic contribution is suppressed 
by $\Lambda_{\rm QCD}^2/m_b^2$ compared to the leading one. 

For $C_0$ we observe that moving from the LO expressions to the NLO expressions does not seem to reduce
the $\mu$ dependence. This has been already noticed in~\cite{Mannel:2023zei,Mannel:2024uar} and is related 
to an accidental cancellation of the large logarithms in the combination of Wilson coefficients in the
leading order contribution to the total rate  
\begin{eqnarray}
	C_{0,\,\rm LO} 
	\propto  (3C_1^2 + 2C_1 C_2 + 3C_2^2)   \, , 
	\label{C0LOex}
\end{eqnarray}
where the leading log $\alpha_s(\mu)  \ln (\mu^2 / M_W^2) $ cancels due to the specific values of the 
anomalous-dimension matrix at order $\alpha_s$. This means in turn that the $\mu$ dependence of the LO
contribution is unnaturally small, and hence no significant reduction of this dependence is observed.

\begin{figure}[ht]
	\centering
		\includegraphics[scale=0.565]{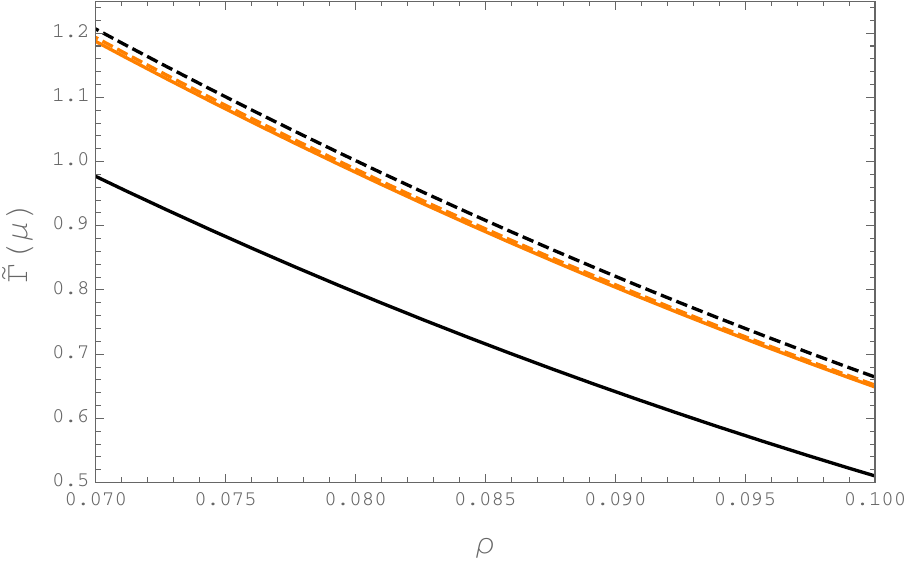}
	\quad
		\includegraphics[scale=0.67]{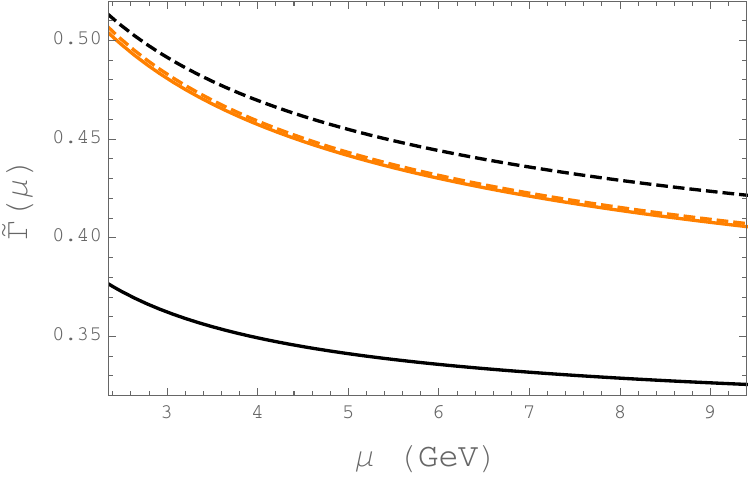}
	\caption{$\tilde{\Gamma}$ as a function of $\rho$ (left) for $\mu = m_b$ and as a function of $\mu$ (right) for the physical value of $\rho$. The black lines show the results at leading power: The solid line is the LO result and the dashed line is the NLO result. The orange lines show the results including the power corrections on top of the leading power at NLO: The solid line includes the LO result and the dashed line the NLO result. Note that the orange lines lie almost on top of each other.}
	\label{fig:GamT}
\end{figure}

Finally, we discuss the impact on the total rate. Let us consider the normalized $b\rightarrow c\bar c s$ decay rate
\begin{equation}
\tilde{\Gamma} =  \frac{\Gamma(b\rightarrow c\bar c s)}{\Gamma_0 |V_{cb}|^2 |V_{cs}|^2}\,,
\label{Gtilde}
\end{equation}
whose dependence on $\rho$ and $\mu$ is shown in fig.~\ref{fig:GamT}.


%

To quantify the size of corrections we insert into Eq.~(\ref{Gtilde}) the representative values of
table~\ref{tab:par} with $\mu = m_b$ and obtain


\begin{eqnarray}
\tilde{\Gamma}
 &=& C_0\bigg(1 - \frac{\mu_\pi^2}{2m_b^2}\bigg)
	+ C_{\mu_G}\frac{\mu_G^2}{2m_b^2}
	\nonumber
  \\
 &=& (0.3473_{\rm LO} + 0.1168_{\rm NLO})\bigg(1 - \frac{\mu_\pi^2}{2m_b^2}\bigg)
	+ (-1.6367_{\rm LO} + 0.2038_{\rm NLO}) \frac{\mu_G^2}{2m_b^2}
	\nonumber
	\\
 &=& (0.3434_{\rm LO} + 0.1155_{\rm NLO})_{0} + (-0.0130_{\rm LO} + 0.0016_{\rm NLO})_{{\mu_G}}
\end{eqnarray}
We observe that the NLO contribution to $C_{\mu_G}$ gives a $-12\%$ correction to the LO $C_{\mu_G}$ coefficient.
Therefore, the new NLO corrections are a small correction to the coefficient,  
in contrast to what it happens in the
$b\rightarrow c\bar u d$ case. 

We also observe that the LO contribution to $C_{\mu_G}$ gives a $-4\%$ corrections to the (leading) partonic term at LO.
The NLO contribution to $C_{\mu_G}$ gives a $0.5\%$ correction to the leading term, in
contrast to the recently computed NNLO corrections to the free quark decay, which give approximately a
$4\%$ correction~\cite{Egner:2024azu}. Therefore, in contrast to what it was observed in the $b\rightarrow c \bar u d$ decay channel~\cite{Mannel:2024uar}, the NNLO corrections to the partonic rate for the $b\rightarrow c \bar c s$ decay channel are more important than the NLO corrections to the chromomagnetic term.
%
%
Overall we conclude that the $\alpha_s$ corrections to the chromomagnetic coefficient in the inclusive nonleptonic $b\rightarrow c\bar c s$ decay leads to small effects.


\section{Conclusions}
\label{sec:conc}
The results presented in this paper complete the calculation of the $1/m_b^2$ suppressed terms in the 
HQE for the total non-leptonic width at order $\alpha_s$. Compared to the previous calculations, where the 
case of masless final state quarks and the case with a single heavy quark in the final state have been
considered, we have now computed the case with two heavy quarks in the final state. 
Technically, all master integrals can be computed analytically and the result can be given in terms
of polylogarithms. The resulting expressions are lengthy, and we provide them as a {\tt Mathematica}
code in the ancillary file \textit{``coefbccs.nb''}.

Overall, the the QCD corrections to the total rates at subleading power $1/m_b^2$ induced by the $b \to c \bar{c} q$ operators in the effective Hamiltonian turn out to be small.
This was expected to some extent, since the available phase space is much smaller than for the massless
case or the case with a single charm quark in the final state.  


With this calculation all ingredients are available for a detailed phenomenological study of
bottom-hadron lifetimes up to $1/m_b^2$, taking into account all (numerically relevant) contributions from the effective weak Hamiltonian,
including also penguin operator effects at leading power. This requires a careful analysis of QCD corrections, including also
a proper definition of the quark masses. This is however beyond the scope of the present paper.

The approach used here could be extended to the calculation of NLO corrections to even higher powers in the $1/m_b$ expansion, e.~g. for the coefficient of the Darwin term.
However this contribution is expected to be even smaller.


\subsection*{Acknowledgments}

We are indebted to the authors of \cite{Egner:2024azu} for providing the NLO master integrals
for the $b\rightarrow c\bar c s$ decay channel. We are specially grateful to Matteo Fael for correspondence regarding
the master integrals and their technical aspects.
DM thanks Sebastian Pögel for discussions on the computation of master integrals by using the
method of differential equations. This research was supported by the Deutsche Forschungsgemeinschaft
(DFG, German Research Foundation) under grant  396021762 - TRR 257
``Particle Physics Phenomenology after the Higgs Discovery''
and has received funding from the European Union’s Horizon 2020
research and innovation program under the Marie Skłodowska-Curie grant
agreement No.~884104 (PSI-FELLOW-III-3i).

\appendix



\section{Coefficients for $b\rightarrow c \bar c s$ channel expanded at $\rho=0$}
\label{app:coefbccs}

In this section we provide analytic results for the perturbative coefficents in the HQE of the inclusive nonleptonic
$b\rightarrow c\bar c s$ decay width. We take $N_c=3$ explicitly for brevity. At LO we provide the exact result whereas at NLO
we provide the first few terms of the expansion around $\rho=0$.
Analytic results in terms of both MPLs and classical polylogarithms with full dependence on $\rho$ and $N_c$ are
provided in the supplemental file \textit{"coefbccs.m"}.

We split the coefficients as follows
\begin{eqnarray}
  C_i &=&
 C_1^2 C_{i,\,11}
 + C_2^2 C_{i,\,22}
 + C_1 C_2 C_{i,\,12}\,,
 \quad\quad (i=0,\,v,\,\mu_G)\,,
 \label{coefsplit1}
\end{eqnarray}
with
\begin{eqnarray}
 C_{i,\,j} &=& C_{i,\,j}^{(0)} + \frac{\alpha_s(\mu)}{\pi}C_{i,\,j}^{(1)}\,,
 \quad\quad (i=0,\,v,\,\mu_G\,;\, j=11,\,22,\,12)\,,
 \label{coefsplit2}
\end{eqnarray}
For the leading power coefficient $C_0$ we obtain
\begin{eqnarray}
C_{0,\,11}^{(0)} &=& C_{0,\,22}^{(0)} = \frac{3}{2}C_{0,\,12}^{(0)} =
-\frac{3 \left(-1+9 x^2+35 x^4+37 x^6-37 x^8-35 x^{10}-9 x^{12}+x^{14}\right)}{\left(1+x^2\right)^7}
\nonumber
\\
&&
\quad\quad\quad\quad\quad\quad\quad\;\;\;
-\frac{144 x^4 \left(1+4
   x^2+5 x^4+4 x^6+x^8\right) \ln (x)}{\left(1+x^2\right)^8}\,,
\\
\nonumber
&&
\\
C_{0,\,11}^{(1)} &=&
\frac{31}{2}-2 \pi ^2
+\rho  \left(32 \left(-10+\pi ^2\right)-96 \ln (\rho )\right)
\nonumber
\\
&&
+\rho ^2 \left(-276+64 \pi ^2+8 \left(21+4 \pi ^2\right) \ln (\rho )-144 \ln ^2(\rho )\right)
\nonumber
\\
&&
+\rho ^3 \left(\frac{16}{3} \left(-97+10 \pi^2\right)-\frac{1600 \ln (\rho )}{3}-32 \ln ^2(\rho )\right)
\nonumber
\\
&&
+\rho ^4 \left(-\frac{7}{2}-108 \pi ^2+\left(\frac{44}{3}-32 \pi ^2\right) \ln (\rho )+316 \ln ^2(\rho )+288 \zeta (3)\right)
+ \mathcal{O}(\rho^{9/2})\,,
\\
\nonumber
&&
\\
C_{0,\,22}^{(1)} &=&
\frac{31}{2}-2 \pi ^2+64 \pi ^2 \rho ^{3/2}-64 \pi ^2 \rho ^{7/2}
+\rho  \left(16 \left(-20+\pi ^2\right)-96 \ln (\rho )\right)
\nonumber
\\
&&
+\rho ^2 \left(-636+16 \pi ^2+8 \left(33+2 \pi ^2\right) \ln (\rho )-144 \ln ^2(\rho )\right)
\nonumber
\\
&&
+\rho^3 \left( -\frac{8}{9}  \left(-337+6 \pi ^2\right)-\frac{2656 \ln (\rho )}{3}-16 \ln ^2(\rho )\right)
\nonumber
\\
&&
+\rho ^4 \left(-\frac{14236}{75}-68 \pi ^2+\left(-\frac{314}{5}+16 \pi ^2\right) \ln (\rho )+360 \ln ^2(\rho )-288
   \zeta (3)\right)
   + \mathcal{O}(\rho^{9/2})\,,
\\
\nonumber
&&
\\
C_{0,\,12}^{(1)} &=&
-17-\frac{4 \pi ^2}{3}-\frac{128}{3} \pi ^2 \rho ^{3/2}+\frac{256}{3} \pi ^2 \rho ^{5/2}-128 \pi ^2 \rho ^{7/2}
-16 \ln \left(\frac{\mu }{m_b}\right)
\nonumber
\\
&&
+\rho  \left(16 \left(21+\pi ^2\right)+256 \ln \left(\frac{\mu}{m_b}\right)-64 \ln (\rho )\right)
\nonumber
\\
&&
+\rho ^2 \left(-328+\left(560+\frac{64 \pi ^2}{3}\right) \ln (\rho )-96 \ln ^2(\rho )+\ln \left(\frac{\mu }{m_b}\right) (-384+384 \ln (\rho ))\right)
\nonumber
\\
&&
+\rho ^3\left(\frac{8}{27} \left(751+66 \pi ^2\right)+512 \ln \left(\frac{\mu }{m_b}\right)-\frac{64}{9} \left(-1+3 \pi ^2\right) \ln (\rho )+16 \ln ^2(\rho )+192 \zeta (3)\right)
\nonumber
\\
&&
+\rho ^4 \bigg(\frac{237341}{675}-\frac{104\pi ^2}{9}+\ln \left(\frac{\mu }{m_b}\right) (-32-384 \ln (\rho ))-\frac{8}{45} \left(2947+120 \pi ^2\right) \ln (\rho )
\nonumber
\\
&&
+\frac{344 \ln ^2(\rho )}{3}+192 \zeta (3)\bigg)
+ \mathcal{O}(\rho^{9/2})\,.
\end{eqnarray}
For the EOM operator coefficient $\bar C_v$ we obtain
\begin{eqnarray}
\bar C_{v,\,11}^{(0)} &=& \bar C_{v,\,22}^{(0)} = \frac{3}{2}\bar C_{v,\,12}^{(0)} =
-\frac{3 \left(-5+13 x^2+63 x^4+9 x^6-9 x^8-63 x^{10}-13 x^{12}+5 x^{14}\right)}{\left(1+x^2\right)^7}
\nonumber
\\
&&
\quad\quad\quad\quad\quad\quad\quad\;\;\;
-\frac{144 x^4 \left(1+4x^2+9 x^4+4 x^6+x^8\right) \ln (x)}{\left(1+x^2\right)^8}\,,
        \\
        &&
        \nonumber
        \\
 \bar C_{v,\,11}^{(1)} &=&
 \frac{65}{6} - 2\pi^2
 +\rho  (72-288 \ln (\rho ))
 \nonumber
\\
&&
 +\rho ^2 \left(-1236+\left(1272-32 \pi ^2\right) \ln (\rho )-144 \ln ^2(\rho )+576 \zeta (3)\right)
 \nonumber
\\
&&
 +\rho ^3 \left(1872-64 \pi ^2+\frac{2560 \ln (\rho )}{3}+32 \ln ^2(\rho )\right)
 \nonumber
\\
&&
 +\rho ^4 \left(\frac{4855}{6}+356 \pi ^2+\left(-1452+96 \pi ^2\right) \ln (\rho )-948 \ln
   ^2(\rho )-864 \zeta (3)\right) + \mathcal{O}(\rho^{9/2})\,,
       \\
        \nonumber
        \\
 \bar C_{v,\,22}^{(1)} &=&
 \frac{65}{6}-2 \pi ^2-32 \pi ^2 \rho ^{3/2}+160 \pi ^2 \rho ^{7/2}
 +\rho  (156-144 \ln (\rho ))
 \nonumber
\\
&&
 +\rho ^2 \left(-468+16 \pi ^2-16 \left(-48+\pi^2\right) \ln (\rho )-48 \ln ^2(\rho )+288 \zeta (3)\right)
 \nonumber
\\
&&
 +\rho ^3 \left(\frac{7172}{9} +8 \pi ^2+888 \ln (\rho )+16 \ln ^2(\rho )\right)
 \nonumber
\\
&&
 +\rho ^4 \left(\frac{32968}{75}+188 \pi ^2+\left(-\frac{3958}{5}-48 \pi ^2\right) \ln (\rho )-1080 \ln ^2(\rho )+864 \zeta (3)\right)
 \nonumber
 \\
 &&
 + \mathcal{O}(\rho^{9/2})\,,
        \\
        \nonumber
        \\
   \bar C_{v,\,12}^{(1)} &=&
   -\frac{1157}{9}-\frac{4 \pi ^2}{3}+\frac{64}{3} \pi ^2 \rho^{3/2}-128 \pi ^2 \rho ^{5/2}+320 \pi ^2 \rho ^{7/2}
   -80 \ln\left(\frac{\mu }{m_b}\right)
   \nonumber
\\
&&
   +\rho  \left( 1112+768 \ln \left(\frac{\mu }{m_b}\right)-288 \ln(\rho )\right)
\nonumber
\\
&&
  +\rho ^2 \bigg( -\frac{16}{3}\left(390+7 \pi ^2\right)+\left(1504-\frac{64 \pi ^2}{3}\right) \ln (\rho )-176 \ln ^2(\rho )
   \nonumber
\\
&&
   +\ln \left(\frac{\mu }{m_b}\right) (-1152+384 \ln (\rho ))+384 \zeta (3)\bigg)
\nonumber
\\
&&
      +\rho ^3 \left(\frac{8408}{27}-\frac{16 \pi ^2}{3}-512 \ln \left(\frac{\mu }{m_b}\right)+\frac{16}{9} \left(-469+24 \pi ^2\right) \ln (\rho )-\frac{64 \ln ^2(\rho )}{3}-384 \zeta (3)\right)
\nonumber
\\
&&
   +\rho ^4 \bigg( \frac{64039}{225}+56 \pi ^2+\left(\frac{16456}{15}+64 \pi ^2\right) \ln (\rho )
   -344 \ln ^2(\rho )+\ln \left(\frac{\mu}{m_b}\right) (864+1152 \ln (\rho ))
\nonumber
\\
&&
   -576 \zeta (3)\bigg) + \mathcal{O}(\rho^{9/2})\,.
\end{eqnarray}
Finally, for the chromomagnetic operator coefficient $C_{\mu_G}$ we obtain
\begin{eqnarray}
  C_{\mu_G,\,11}^{(0)} &=& C_{\mu_G,\,22}^{(0)} =
  \frac{3 \left(-3-5 x^2-7 x^4-65 x^6+65 x^8+7 x^{10}+5 x^{12}+3 x^{14}\right)}{\left(1+x^2\right)^7}
   \nonumber
 \\
 &&
 \quad\quad\quad\quad
  -\frac{144 x^4 \left(1+4x^2+x^4+4 x^6+x^8\right) \ln (x)}{\left(1+x^2\right)^8}\,,
 \\
 &&
 \nonumber
 \\
 C_{\mu_G,\,12}^{(0)} &=&
 \frac{2 \left(-19-93 x^2-223 x^4-209 x^6+209 x^8+223 x^{10}+93 x^{12}+19 x^{14}\right)}{\left(1+x^2\right)^7}
 \nonumber
 \\
 &&
 -\frac{96 x^2\left(2+13 x^2+30 x^4+33 x^6+30 x^8+13 x^{10}+2 x^{12}\right) \ln (x)}{\left(1+x^2\right)^8}\,,
 \\
 &&
 \nonumber
 \\
C_{\mu_G,\,11}^{(1)} &=&
\frac{1}{18} \left(811+6 \pi ^2\right)-\frac{80 \pi ^2 \sqrt{\rho }}{3}-\frac{496}{9} \pi ^2 \rho ^{3/2}-48 \pi ^2 \rho ^{5/2}+\frac{208}{5} \pi ^2 \rho ^{7/2}+48 \ln \left(\frac{\mu }{m_b}\right)
\nonumber
\\
&&
+\rho
   \left(49+\frac{244 \pi ^2}{3}+8 \left(17+\pi ^2\right) \ln (\rho )+10 \ln ^2(\rho )+\ln \left(\frac{\mu }{m_b}\right) (-72+144 \ln (\rho ))\right)
  \nonumber
  \\
  &&
   +\rho ^2 \left(\frac{38515}{54}+\frac{59 \pi^2}{3}+288 \ln \left(\frac{\mu }{m_b}\right)+\left(-\frac{1613}{3}+52 \pi ^2\right) \ln (\rho )-\frac{446 \ln ^2(\rho )}{3}-180 \zeta (3)\right)
   \nonumber
\\
&&
   +\rho ^3 \bigg(-\frac{3607621}{1620}+\frac{4838 \pi ^2}{27}+\ln
   \left(\frac{\mu }{m_b}\right) (-96-288 \ln (\rho ))
      \nonumber
\\
&&
   +\left(-\frac{48557}{27}+\frac{448 \pi ^2}{9}\right) \ln (\rho )-\frac{2468 \ln ^2(\rho )}{9}-448 \zeta (3)\bigg)
   +\rho ^4 \bigg(-\frac{5618821}{22050}
\nonumber
\\
&&
   -402 \pi ^2-480 \ln
   \left(\frac{\mu }{m_b}\right)+\left(\frac{25663}{35}-160 \pi ^2\right) \ln (\rho )+1395 \ln ^2(\rho )
   +1440 \zeta (3)\bigg)
    +\mathcal{O}(\rho^{9/2})\,,
    \nonumber
    \\
    &&
\end{eqnarray}

\begin{eqnarray}
C_{\mu_G,\,22}^{(1)} &=&
\frac{1}{18} \left(811+6 \pi ^2\right)-\frac{40 \pi ^2 \sqrt{\rho }}{3}-\frac{416}{9} \pi ^2 \rho ^{3/2}+24 \pi ^2 \rho ^{5/2}-\frac{1040}{3} \pi ^2 \rho ^{7/2}+48 \ln \left(\frac{\mu }{m_b}\right)
\nonumber
\\
&&
+\rho
   \left(46+\frac{134 \pi ^2}{3}+\left(131+8 \pi ^2\right) \ln (\rho )+5 \ln ^2(\rho )+\ln \left(\frac{\mu }{m_b}\right) (-72+144 \ln (\rho ))\right)
   \nonumber
\\
&&
   +\rho ^2 \left(\frac{101}{27} \left(29+18 \pi ^2\right)+288 \ln \left(\frac{\mu}{m_b}\right) + \frac{1}{3}\left(-263+80 \pi ^2\right) \ln (\rho )-\frac{259 \ln ^2(\rho )}{3}-96 \zeta (3)\right)
      \nonumber
\\
&&
   +\rho ^3 \bigg(-\frac{1498501}{3240}-\frac{995 \pi ^2}{27}+\ln \left(\frac{\mu }{m_b}\right)
   (-96-288 \ln (\rho ))-\frac{23}{54} \left(4897+48 \pi ^2\right) \ln (\rho )
\nonumber
\\
&&
   -\frac{3322 \ln ^2(\rho )}{9}+256 \zeta (3)\bigg)
   +\rho ^4 \bigg( \frac{704817}{4900}-\frac{544 \pi ^2}{3}-480 \ln
   \left(\frac{\mu }{m_b}\right)
   \nonumber
\\
&&
   +\left(-\frac{134117}{210}+80 \pi ^2\right) \ln (\rho )+1439 \ln ^2(\rho )-1440 \zeta (3)\bigg)
   +\mathcal{O}(\rho^{9/2})\,,
\end{eqnarray}

\begin{eqnarray}
 C_{\mu_G,\,12}^{(1)} &=&
 \frac{1}{27} \left(-1751+86 \pi ^2\right)-\frac{560 \pi ^2 \sqrt{\rho }}{9}-\frac{8416}{27} \pi ^2 \rho ^{3/2}-112 \pi ^2 \rho ^{5/2}+\frac{1216}{15} \pi ^2 \rho ^{7/2}+16 \ln \left(\frac{\mu }{m_b}\right)
 \nonumber
\\
&&
   +\rho  \bigg(\frac{2764}{3}+254 \pi ^2+\ln \left(\frac{\mu }{m_b}\right) (-208-96 \ln (\rho ))+\frac{2}{3} \left(215+56 \pi ^2\right) \ln (\rho )+\frac{184 \ln ^2(\rho )}{3}
    \nonumber
\\
&&
   -112 \zeta (3)\bigg)
 +\rho ^2
   \bigg(\frac{55715}{81}-\frac{362 \pi ^2}{9}+\frac{2}{9} \left(-7069+236 \pi ^2\right) \ln (\rho )-\frac{1136 \ln ^2(\rho )}{9}
\nonumber
\\
&&
   +\ln \left(\frac{\mu }{m_b}\right) (192+384 \ln (\rho ))-280 \zeta (3)\bigg)
\nonumber
\\
&&
   +\rho ^3
   \bigg(\frac{6872509}{2430}-\frac{4286 \pi ^2}{27}+\left(\frac{28690}{81}+\frac{32 \pi ^2}{9}\right) \ln (\rho )+\frac{12436 \ln^2(\rho )}{27}
   \nonumber
\\
&&
   +\ln \left(\frac{\mu }{m_b}\right) (1600+192 \ln (\rho ))-192 \zeta(3)\bigg)
\nonumber
\\
&&
   +\rho ^4
   \bigg(-\frac{9088694}{4725}-\frac{212 \pi ^2}{3}+\ln \left(\frac{\mu }{m_b}\right) (-608-1920 \ln (\rho ))-\frac{2}{45} \left(7369+2400 \pi ^2\right) \ln (\rho )
\nonumber
\\
&&
   +\frac{746 \ln ^2(\rho )}{3}+960 \zeta (3)\bigg)
   +\mathcal{O}(\rho^{9/2})\,.
\end{eqnarray}


%

\bibliographystyle{JHEP}
\bibliography{NLbccs_240325}

\providecommand{\href}[2]{#2}\begingroup\raggedright\begin{thebibliography}{10}

\bibitem{Chay:1990da}
J.~Chay, H.~Georgi and B.~Grinstein, \emph{{Lepton energy distributions in
  heavy meson decays from QCD}},
  \href{https://doi.org/10.1016/0370-2693(90)90916-T}{\emph{Phys. Lett. B}
  {\bfseries 247} (1990) 399}.

\bibitem{Bigi:1992su}
I.I.Y.~Bigi, N.G.~Uraltsev and A.I.~Vainshtein, \emph{{Nonperturbative
  corrections to inclusive beauty and charm decays: QCD versus phenomenological
  models}}, \href{https://doi.org/10.1016/0370-2693(92)90908-M}{\emph{Phys.
  Lett. B} {\bfseries 293} (1992) 430}
  [\href{https://arxiv.org/abs/hep-ph/9207214}{{\ttfamily hep-ph/9207214}}].

\bibitem{Bigi:1993fe}
I.I.Y.~Bigi, M.A.~Shifman, N.G.~Uraltsev and A.I.~Vainshtein, \emph{{QCD
  predictions for lepton spectra in inclusive heavy flavor decays}},
  \href{https://doi.org/10.1103/PhysRevLett.71.496}{\emph{Phys. Rev. Lett.}
  {\bfseries 71} (1993) 496}
  [\href{https://arxiv.org/abs/hep-ph/9304225}{{\ttfamily hep-ph/9304225}}].

\bibitem{Blok:1993va}
B.~Blok, L.~Koyrakh, M.A.~Shifman and A.I.~Vainshtein, \emph{{Differential
  distributions in semileptonic decays of the heavy flavors in QCD}},
  \href{https://doi.org/10.1103/PhysRevD.50.3572}{\emph{Phys. Rev. D}
  {\bfseries 49} (1994) 3356}
  [\href{https://arxiv.org/abs/hep-ph/9307247}{{\ttfamily hep-ph/9307247}}].

\bibitem{Manohar:1993qn}
A.V.~Manohar and M.B.~Wise, \emph{{Inclusive semileptonic B and polarized
  Lambda(b) decays from QCD}},
  \href{https://doi.org/10.1103/PhysRevD.49.1310}{\emph{Phys. Rev. D}
  {\bfseries 49} (1994) 1310}
  [\href{https://arxiv.org/abs/hep-ph/9308246}{{\ttfamily hep-ph/9308246}}].

\bibitem{Lenz:2014jha}
A.~Lenz, \emph{{Lifetimes and heavy quark expansion}},
  \href{https://doi.org/10.1142/S0217751X15430058}{\emph{Int. J. Mod. Phys. A}
  {\bfseries 30} (2015) 1543005}
  [\href{https://arxiv.org/abs/1405.3601}{{\ttfamily 1405.3601}}].

\bibitem{Lenz:2022rbq}
A.~Lenz, M.L.~Piscopo and A.V.~Rusov, \emph{{Disintegration of beauty: a
  precision study}}, \href{https://doi.org/10.1007/JHEP01(2023)004}{\emph{JHEP}
  {\bfseries 01} (2023) 004}
  [\href{https://arxiv.org/abs/2208.02643}{{\ttfamily 2208.02643}}].

\bibitem{Gratrex:2023pfn}
J.~Gratrex, A.~Lenz, B.~Meli\'c, I.~Ni\v{s}and\v{z}i\'c, M.L.~Piscopo and
  A.V.~Rusov, \emph{{Quark-hadron duality at work: lifetimes of bottom
  baryons}}, \href{https://doi.org/10.1007/JHEP04(2023)034}{\emph{JHEP}
  {\bfseries 04} (2023) 034}
  [\href{https://arxiv.org/abs/2301.07698}{{\ttfamily 2301.07698}}].

\bibitem{Albrecht:2024oyn}
J.~Albrecht, F.~Bernlochner, A.~Lenz and A.~Rusov, \emph{{Lifetimes of
  b-hadrons and mixing of neutral B-mesons: theoretical and experimental
  status}}, \href{https://doi.org/10.1140/epjs/s11734-024-01124-3}{\emph{Eur.
  Phys. J. ST} {\bfseries 233} (2024) 359}
  [\href{https://arxiv.org/abs/2402.04224}{{\ttfamily 2402.04224}}].

\bibitem{King:2021xqp}
D.~King, A.~Lenz, M.L.~Piscopo, T.~Rauh, A.V.~Rusov and C.~Vlahos,
  \emph{{Revisiting inclusive decay widths of charmed mesons}},
  \href{https://doi.org/10.1007/JHEP08(2022)241}{\emph{JHEP} {\bfseries 08}
  (2022) 241} [\href{https://arxiv.org/abs/2109.13219}{{\ttfamily
  2109.13219}}].

\bibitem{Gratrex:2022xpm}
J.~Gratrex, B.~Meli\'c and I.~Ni\v{s}and\v{z}i\'c, \emph{{Lifetimes of singly
  charmed hadrons}}, \href{https://doi.org/10.1007/JHEP07(2022)058}{\emph{JHEP}
  {\bfseries 07} (2022) 058}
  [\href{https://arxiv.org/abs/2204.11935}{{\ttfamily 2204.11935}}].

\bibitem{Dulibic:2023jeu}
L.~Dulibi\'c, J.~Gratrex, B.~Meli\'c and I.~Ni\v{s}and\v{z}i\'c,
  \emph{{Revisiting lifetimes of doubly charmed baryons}},
  \href{https://doi.org/10.1007/JHEP07(2023)061}{\emph{JHEP} {\bfseries 07}
  (2023) 061} [\href{https://arxiv.org/abs/2305.02243}{{\ttfamily
  2305.02243}}].

\bibitem{HFLAV:2022esi}
{\scshape HFLAV} collaboration, \emph{{Averages of b-hadron, c-hadron, and
  \ensuremath{\tau}-lepton properties as of 2021}},
  \href{https://doi.org/10.1103/PhysRevD.107.052008}{\emph{Phys. Rev. D}
  {\bfseries 107} (2023) 052008}
  [\href{https://arxiv.org/abs/2206.07501}{{\ttfamily 2206.07501}}].

\bibitem{Altarelli:1980fi}
G.~Altarelli, G.~Curci, G.~Martinelli and S.~Petrarca, \emph{{QCD Nonleading
  Corrections to Weak Decays as an Application of Regularization by Dimensional
  Reduction}}, \href{https://doi.org/10.1016/0550-3213(81)90473-9}{\emph{Nucl.
  Phys. B} {\bfseries 187} (1981) 461}.

\bibitem{Buchalla:1992gc}
G.~Buchalla, \emph{{O (alpha-s) QCD corrections to charm quark decay in
  dimensional regularization with nonanticommuting gamma-5}},
  \href{https://doi.org/10.1016/0550-3213(93)90081-Y}{\emph{Nucl. Phys. B}
  {\bfseries 391} (1993) 501}.

\bibitem{Bagan:1994zd}
E.~Bagan, P.~Ball, V.M.~Braun and P.~Gosdzinsky, \emph{{Charm quark mass
  dependence of QCD corrections to nonleptonic inclusive B decays}},
  \href{https://doi.org/10.1016/0550-3213(94)90591-6}{\emph{Nucl. Phys. B}
  {\bfseries 432} (1994) 3}
  [\href{https://arxiv.org/abs/hep-ph/9408306}{{\ttfamily hep-ph/9408306}}].

\bibitem{Czarnecki:2005vr}
A.~Czarnecki, M.~Slusarczyk and F.V.~Tkachov, \emph{{Enhancement of the
  hadronic b quark decays}},
  \href{https://doi.org/10.1103/PhysRevLett.96.171803}{\emph{Phys. Rev. Lett.}
  {\bfseries 96} (2006) 171803}
  [\href{https://arxiv.org/abs/hep-ph/0511004}{{\ttfamily hep-ph/0511004}}].

\bibitem{Krinner:2013cja}
F.~Krinner, A.~Lenz and T.~Rauh, \emph{{The inclusive decay $b \to c\bar{c}s$
  revisited}},
  \href{https://doi.org/10.1016/j.nuclphysb.2013.07.028}{\emph{Nucl. Phys. B}
  {\bfseries 876} (2013) 31} [\href{https://arxiv.org/abs/1305.5390}{{\ttfamily
  1305.5390}}].

\bibitem{Egner:2024azu}
M.~Egner, M.~Fael, K.~Sch\"onwald and M.~Steinhauser, \emph{{Nonleptonic
  $B$-meson decays to next-to-next-to-leading order}},
  \href{https://arxiv.org/abs/2406.19456}{{\ttfamily 2406.19456}}.

\bibitem{Egner:2024lay}
M.~Egner, M.~Fael, A.~Lenz, M.L.~Piscopo, A.V.~Rusov, K.~Sch\"onwald et~al.,
  \emph{{Total decay rates of $B$ mesons at NNLO-QCD}},
  \href{https://arxiv.org/abs/2412.14035}{{\ttfamily 2412.14035}}.

\bibitem{Blok:1992hw}
B.~Blok and M.A.~Shifman, \emph{{The Rule of discarding 1/N(c) in inclusive
  weak decays. 1.}},
  \href{https://doi.org/10.1016/0550-3213(93)90504-I}{\emph{Nucl. Phys. B}
  {\bfseries 399} (1993) 441}
  [\href{https://arxiv.org/abs/hep-ph/9207236}{{\ttfamily hep-ph/9207236}}].

\bibitem{Blok:1992he}
B.~Blok and M.A.~Shifman, \emph{{The Rule of discarding 1/N(c) in inclusive
  weak decays. 2.}},
  \href{https://doi.org/10.1016/0550-3213(93)90505-J}{\emph{Nucl. Phys. B}
  {\bfseries 399} (1993) 459}
  [\href{https://arxiv.org/abs/hep-ph/9209289}{{\ttfamily hep-ph/9209289}}].

\bibitem{Mannel:2023zei}
T.~Mannel, D.~Moreno and A.A.~Pivovarov, \emph{{Heavy-quark expansion for
  lifetimes: Toward the QCD corrections to power suppressed terms}},
  \href{https://doi.org/10.1103/PhysRevD.107.114026}{\emph{Phys. Rev. D}
  {\bfseries 107} (2023) 114026}
  [\href{https://arxiv.org/abs/2304.08964}{{\ttfamily 2304.08964}}].

\bibitem{Lenz:2020oce}
A.~Lenz, M.L.~Piscopo and A.V.~Rusov, \emph{{Contribution of the Darwin
  operator to non-leptonic decays of heavy quarks}},
  \href{https://doi.org/10.1007/JHEP12(2020)199}{\emph{JHEP} {\bfseries 12}
  (2020) 199} [\href{https://arxiv.org/abs/2004.09527}{{\ttfamily
  2004.09527}}].

\bibitem{Mannel:2020fts}
T.~Mannel, D.~Moreno and A.~Pivovarov, \emph{{Heavy quark expansion for heavy
  hadron lifetimes: completing the $ 1/{m}_b^3 $ corrections}},
  \href{https://doi.org/10.1007/JHEP08(2020)089}{\emph{JHEP} {\bfseries 08}
  (2020) 089} [\href{https://arxiv.org/abs/2004.09485}{{\ttfamily
  2004.09485}}].

\bibitem{Moreno:2020rmk}
D.~Moreno, \emph{{Completing $1/m_b^3$ corrections to non-leptonic
  bottom-to-up-quark decays}},
  \href{https://doi.org/10.1007/JHEP01(2021)051}{\emph{JHEP} {\bfseries 01}
  (2021) 051} [\href{https://arxiv.org/abs/2009.08756}{{\ttfamily
  2009.08756}}].

\bibitem{Beneke:2002rj}
M.~Beneke, G.~Buchalla, C.~Greub, A.~Lenz and U.~Nierste, \emph{{The $B^+
  -B^0_d$ Lifetime Difference Beyond Leading Logarithms}},
  \href{https://doi.org/10.1016/S0550-3213(02)00561-8}{\emph{Nucl. Phys. B}
  {\bfseries 639} (2002) 389}
  [\href{https://arxiv.org/abs/hep-ph/0202106}{{\ttfamily hep-ph/0202106}}].

\bibitem{Franco:2002fc}
E.~Franco, V.~Lubicz, F.~Mescia and C.~Tarantino, \emph{{Lifetime ratios of
  beauty hadrons at the next-to-leading order in QCD}},
  \href{https://doi.org/10.1016/S0550-3213(02)00262-6}{\emph{Nucl. Phys. B}
  {\bfseries 633} (2002) 212}
  [\href{https://arxiv.org/abs/hep-ph/0203089}{{\ttfamily hep-ph/0203089}}].

\bibitem{Gabbiani:2004tp}
F.~Gabbiani, A.I.~Onishchenko and A.A.~Petrov, \emph{{Spectator effects and
  lifetimes of heavy hadrons}},
  \href{https://doi.org/10.1103/PhysRevD.70.094031}{\emph{Phys. Rev. D}
  {\bfseries 70} (2004) 094031}
  [\href{https://arxiv.org/abs/hep-ph/0407004}{{\ttfamily hep-ph/0407004}}].

\bibitem{Mannel:2024uar}
T.~Mannel, D.~Moreno and A.A.~Pivovarov, \emph{{QCD corrections at subleading
  power for inclusive nonleptonic b\textrightarrow{}cu\textasciimacron{}d
  decays}}, \href{https://doi.org/10.1103/PhysRevD.110.094011}{\emph{Phys. Rev.
  D} {\bfseries 110} (2024) 094011}
  [\href{https://arxiv.org/abs/2408.06767}{{\ttfamily 2408.06767}}].

\bibitem{goncharov2011multiplepolylogarithmscyclotomymodular}
A.B.~Goncharov, \emph{Multiple polylogarithms, cyclotomy and modular
  complexes},  2011.

\bibitem{goncharov2001multiplepolylogarithmsmixedtate}
A.B.~Goncharov, \emph{Multiple polylogarithms and mixed tate motives},  2001.

\bibitem{Buras:1989xd}
A.J.~Buras and P.H.~Weisz, \emph{{QCD Nonleading Corrections to Weak Decays in
  Dimensional Regularization and 't Hooft-Veltman Schemes}},
  \href{https://doi.org/10.1016/0550-3213(90)90223-Z}{\emph{Nucl. Phys. B}
  {\bfseries 333} (1990) 66}.

\bibitem{Dugan:1990df}
M.J.~Dugan and B.~Grinstein, \emph{{On the vanishing of evanescent operators}},
  \href{https://doi.org/10.1016/0370-2693(91)90680-O}{\emph{Phys. Lett. B}
  {\bfseries 256} (1991) 239}.

\bibitem{Herrlich:1994kh}
S.~Herrlich and U.~Nierste, \emph{{Evanescent operators, scheme dependences and
  double insertions}},
  \href{https://doi.org/10.1016/0550-3213(95)00474-7}{\emph{Nucl. Phys. B}
  {\bfseries 455} (1995) 39}
  [\href{https://arxiv.org/abs/hep-ph/9412375}{{\ttfamily hep-ph/9412375}}].

\bibitem{Gorbahn:2004my}
M.~Gorbahn and U.~Haisch, \emph{{Effective Hamiltonian for non-leptonic
  $|\Delta F| = 1$ decays at NNLO in QCD}},
  \href{https://doi.org/10.1016/j.nuclphysb.2005.01.047}{\emph{Nucl. Phys. B}
  {\bfseries 713} (2005) 291}
  [\href{https://arxiv.org/abs/hep-ph/0411071}{{\ttfamily hep-ph/0411071}}].

\bibitem{Mannel:2021zzr}
T.~Mannel, D.~Moreno and A.A.~Pivovarov, \emph{{NLO QCD corrections to
  inclusive $b \rightarrow c \ell \bar{\nu}$decay spectra up to~$1/m_Q^3$}},
  \href{https://doi.org/10.1103/PhysRevD.105.054033}{\emph{Phys. Rev. D}
  {\bfseries 105} (2022) 054033}
  [\href{https://arxiv.org/abs/2112.03875}{{\ttfamily 2112.03875}}].

\bibitem{Moreno:2022goo}
D.~Moreno, \emph{{NLO QCD corrections to inclusive semitauonic weak decays of
  heavy hadrons up to 1/mb3}},
  \href{https://doi.org/10.1103/PhysRevD.106.114008}{\emph{Phys. Rev. D}
  {\bfseries 106} (2022) 114008}
  [\href{https://arxiv.org/abs/2207.14245}{{\ttfamily 2207.14245}}].

\bibitem{Moreno:2024bgq}
D.~Moreno, \emph{{QCD corrections to the Darwin coefficient in inclusive
  semileptonic
  B\textrightarrow{}Xu\ensuremath{\ell}\ensuremath{\nu}\textasciimacron{}\ensuremath{\ell}
  decays}}, \href{https://doi.org/10.1103/PhysRevD.109.074030}{\emph{Phys. Rev.
  D} {\bfseries 109} (2024) 074030}
  [\href{https://arxiv.org/abs/2402.13805}{{\ttfamily 2402.13805}}].

\bibitem{Mannel:2018mqv}
T.~Mannel and K.K.~Vos, \emph{{Reparametrization Invariance and Partial
  Re-Summations of the Heavy Quark Expansion}},
  \href{https://doi.org/10.1007/JHEP06(2018)115}{\emph{JHEP} {\bfseries 06}
  (2018) 115} [\href{https://arxiv.org/abs/1802.09409}{{\ttfamily
  1802.09409}}].

\bibitem{Jamin:1991dp}
M.~Jamin and M.E.~Lautenbacher, \emph{{TRACER: Version 1.1: A Mathematica
  package for gamma algebra in arbitrary dimensions}},
  \href{https://doi.org/10.1016/0010-4655(93)90097-V}{\emph{Comput. Phys.
  Commun.} {\bfseries 74} (1993) 265}.

\bibitem{Sjodahl:2012nk}
M.~Sj\"odahl, \emph{{ColorMath - A package for color summed calculations in
  SU(Nc)}}, \href{https://doi.org/10.1140/epjc/s10052-013-2310-4}{\emph{Eur.
  Phys. J. C} {\bfseries 73} (2013) 2310}
  [\href{https://arxiv.org/abs/1211.2099}{{\ttfamily 1211.2099}}].

\bibitem{Lee:2012cn}
R.N.~Lee, \emph{{Presenting LiteRed: a tool for the Loop InTEgrals REDuction}},
   \href{https://arxiv.org/abs/1212.2685}{{\ttfamily 1212.2685}}.

\bibitem{Lee:2013mka}
R.N.~Lee, \emph{{LiteRed 1.4: a powerful tool for reduction of multiloop
  integrals}}, \href{https://doi.org/10.1088/1742-6596/523/1/012059}{\emph{J.
  Phys. Conf. Ser.} {\bfseries 523} (2014) 012059}
  [\href{https://arxiv.org/abs/1310.1145}{{\ttfamily 1310.1145}}].

\bibitem{Duhr:2019tlz}
C.~Duhr and F.~Dulat, \emph{{PolyLogTools \textemdash{} polylogs for the
  masses}}, \href{https://doi.org/10.1007/JHEP08(2019)135}{\emph{JHEP}
  {\bfseries 08} (2019) 135}
  [\href{https://arxiv.org/abs/1904.07279}{{\ttfamily 1904.07279}}].

\bibitem{Maitre:2005uu}
D.~Maitre, \emph{{HPL, a mathematica implementation of the harmonic
  polylogarithms}},
  \href{https://doi.org/10.1016/j.cpc.2005.10.008}{\emph{Comput. Phys. Commun.}
  {\bfseries 174} (2006) 222}
  [\href{https://arxiv.org/abs/hep-ph/0507152}{{\ttfamily hep-ph/0507152}}].

\end{thebibliography}\endgroup

\end{document}